\documentclass{elsart}

\usepackage{graphicx} 
\usepackage{dcolumn}  

\graphicspath{{epss/}{/data/gershon/analysis/dpart_fits/epss/}}

\def\cal{\mathcal}

\newcommand{\Rdp}   {R_{D^*\pi}}

\newcommand{\ddp}   {\delta_{D^*\pi}}

\newcommand{\jpsi} {J/ \psi}

\newcommand{\cts}   {\cos \theta_{\rm hel}}
\newcommand{\ctssq} {\cos \theta_{\rm hel}^{2}}
\newcommand{\cdfs}  {\cos \delta_{\pi_f \pi_s}}
\newcommand{\cdfl}  {\cos \delta_{\pi_f l}}

\begin{document}

\def\myspecial#1{}                   

\begin{frontmatter}

\vspace*{-3\baselineskip}
\begin{flushleft}
 \resizebox{!}{28mm}{\includegraphics{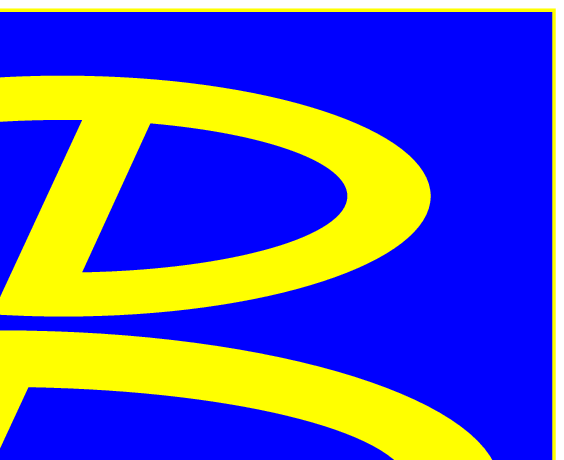}}
\end{flushleft}
\vspace*{-28mm}
\begin{flushright}
  Belle Preprint 2005-8 \\
  KEK Preprint 2004-104
\end{flushright}

\vspace*{12mm}

\title{ 
  \boldmath
  Time-Dependent $CP$ Violation Effects in 
  Partially Reconstructed $B^0 \to D^{*\mp} \pi^\pm$ Decays
}

\collab{Belle Collaboration}
   \author[KEK]{T.~Gershon}, 
   \author[KEK]{K.~Abe}, 
   \author[TohokuGakuin]{K.~Abe}, 
   \author[KEK]{I.~Adachi}, 
   \author[Tokyo]{H.~Aihara}, 
   \author[Tsukuba]{Y.~Asano}, 
   \author[ITEP]{T.~Aushev}, 
   \author[Cincinnati]{S.~Bahinipati}, 
   \author[Sydney]{A.~M.~Bakich}, 
   \author[Tata]{S.~Banerjee}, 
% ...

% ...
\address[Cincinnati]{University of Cincinnati, Cincinnati, OH, USA}
% ...
\address[KEK]{High Energy Accelerator Research Organization (KEK), Tsukuba, Japan}
% ...
\address[ITEP]{Institute for Theoretical and Experimental Physics, Moscow, Russia}
% ...
\address[Sydney]{University of Sydney, Sydney, NSW, Australia}
\address[Tata]{Tata Institute of Fundamental Research, Bombay, India}
% ...
\address[TohokuGakuin]{Tohoku Gakuin University, Tagajo, Japan}
% ...
\address[Tokyo]{Department of Physics, University of Tokyo, Tokyo, Japan}
% ...
\address[Tsukuba]{University of Tsukuba, Tsukuba, Japan}
% ...
\thanks[NovaGorica]{on leave from Nova Gorica Polytechnic, Nova Gorica, Slovenia}

\myspecial{!userdict begin /bop-hook{gsave 280 50 translate 0 rotate
    /Times-Roman findfont 18 scalefont setfont
    0 0 moveto 0.70 setgray
    (\mySpecialText)
    show grestore}def end}

\begin{abstract}

We report measurements of time-dependent decay rates for $B^0 \to D^{*\mp} \pi^\pm$ 
decays and extraction of $CP$ violation parameters related to $\phi_3$. 
We use a partial reconstruction technique, whereby signal events are 
identified using information only from the primary pion and the 
charged pion from the decay of the $D^{*\mp}$. 
The analysis uses $140 \ {\rm fb}^{-1}$ of data accumulated at the 
$\Upsilon(4S)$ resonance with the Belle detector at the KEKB 
asymmetric-energy $e^{+}e^{-}$ collider.
We measure the $CP$ violation parameters
$S^+ = 0.035 \pm 0.041 \ ({\rm stat}) \pm 0.018 \ ({\rm syst})$ and
$S^- = 0.025 \pm 0.041 \ ({\rm stat}) \pm 0.018 \ ({\rm syst})$.

\end{abstract}

\end{frontmatter}

{\renewcommand{\thefootnote}{\fnsymbol{footnote}}}
\setcounter{footnote}{0}

\begin{figure}[!tbh]
  \begin{center}
    \includegraphics[width=0.24\textwidth, bb= 0 0 308 130]{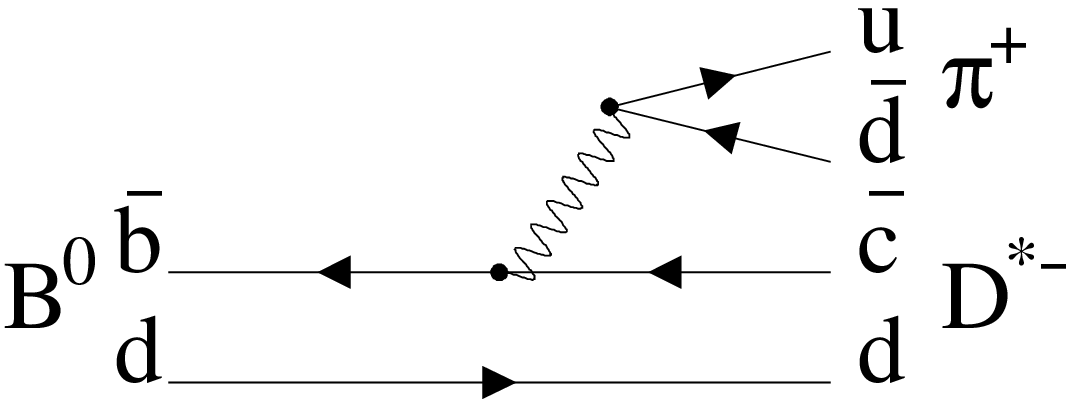}
    \hspace{5mm}
    \includegraphics[width=0.24\textwidth]{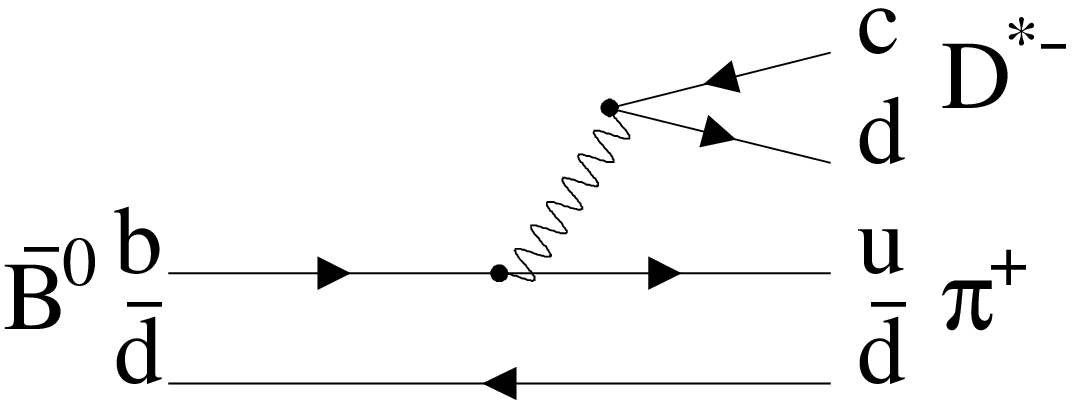}
    \caption{
      \label{fig:feynman}
      Diagrams for 
      (left) $B^0 \to D^{*-}\pi^+$ and 
      (right) $\bar{B}^0 \to D^{*-}\pi^+$.
      Those for $\bar{B}^0 \to D^{*+}\pi^-$ and $B^0 \to D^{*+}\pi^-$
      can be obtained by charge conjugation.
    }
  \end{center}
\end{figure}

Within the Standard Model (SM), 
$CP$ violation arises due to a single phase
in the Cabibbo-Kobayashi-Maskawa (CKM) quark mixing matrix~\cite{KM}.
Precise measurements of CKM matrix parameters therefore
constrain the SM, and may reveal new sources of $CP$ violation.
Measurements of the time-dependent decay rates of $B^0 \to D^{*\mp}\pi^{\pm}$
provide a theoretically clean method for extracting 
$\sin(2\phi_1+\phi_3)$~\cite{dunietz}.
As shown in Fig.~\ref{fig:feynman},
these decays can be mediated by both
Cabibbo-favoured ($V_{cb}^*V_{ud}$) and 
Cabibbo-suppressed ($V_{ub}^*V_{cd}$) amplitudes,
which have a relative weak phase $\phi_3$.

The time-dependent decay rates are given by~\cite{fleischer}
\begin{equation}
  \begin{array}{ccc}
    \label{eq:pdf}
    P(B^{0} \to D^{*+} \pi^-) & = & 
    \frac{1}{8\tau_{B^0}} e^{- \left| \Delta t \right|/\tau_{B^0}}
    \left[
      1 - C \cos (\Delta m \Delta t) - S^+ \sin (\Delta m \Delta t) 
    \right],  
    \\[1.2ex]
    P(B^{0} \to D^{*-} \pi^+) & = &
    \frac{1}{8\tau_{B^0}} e^{- \left| \Delta t \right|/\tau_{B^0}}
    \left[
      1 + C \cos (\Delta m \Delta t) - S^- \sin (\Delta m \Delta t) 
    \right],  
    \\[1.2ex]
    P(\bar{B}^{0} \to D^{*+} \pi^-) & = & 
    \frac{1}{8\tau_{B^0}} e^{- \left| \Delta t \right|/\tau_{B^0}}
    \left[
      1 + C \cos (\Delta m \Delta t) + S^+ \sin (\Delta m \Delta t) 
    \right],  
    \\[1.2ex]
    P(\bar{B}^{0} \to D^{*-} \pi^+) & = &
    \frac{1}{8\tau_{B^0}} e^{- \left| \Delta t \right|/\tau_{B^0}}
    \left[
      1 - C \cos (\Delta m \Delta t) + S^- \sin (\Delta m \Delta t) 
    \right],  
  \end{array}
\end{equation}
where $\Delta t$ is the difference between the time of the decay
and the time that the flavour of the $B$ meson is tagged,
$\tau_{B^0}$ is the average neutral $B$ meson lifetime, 
$\Delta m$ is the $B^0$\textemdash$\bar{B}^0$ mixing parameter,
$C = \left( 1 - \Rdp^2 \right) / \left( 1 + \Rdp^2 \right)$ and
$S^{\pm} = - 2 \Rdp \sin(2\phi_1+\phi_3 \pm \ddp) / \left( 1 + \Rdp^2 \right)$.
The parameter $\Rdp$ is the ratio of the magnitudes of the 
suppressed and favoured amplitudes,
and $\ddp$ is their strong phase difference.
The factor of $-1$ in the relation between $S^{\pm}$ and $\sin(2\phi_1+\phi_3 \pm \ddp)$
arises due to the angular momentum of the final state (which is 1 for $D^* \pi$);
note that it may equivalently be absorbed in a redefinition $\ddp \to \ddp + \pi$.
The value of $\Rdp$ is predicted to be about $0.02$~\cite{csr},
but is not yet measured.
Therefore, 
we neglect terms of ${\cal O}\left( R^2 \right)$ (and hence take $C = 1$),
and do not attempt to extract $\sin(2\phi_1+\phi_3)$, but simply measure $S^\pm$.

In order to obtain a large event sample, 
necessary to probe the small $CP$ violating effect in these decays,
we employ a partial reconstruction technique~\cite{zheng}.
The signal is distinguished from background on the basis of 
the kinematics of the ``fast'' pion, from the decay $B \to D^* \pi_f$,
and the ``slow'' pion, from the decay $D^* \to D \pi_s$, alone;
no attempt is made to reconstruct the  $D$ meson from its decay products.
Background from continuum $e^+e^- \to q\bar{q} \ (q = u,d,s,c)$
events is dramatically reduced by requiring the presence 
of a high-momentum lepton in the event,
which also serves to tag the flavour of the associated $B$ in the event.
Since semileptonic $B$ decays are flavour-specific,
there is no possibility of $CP$ violation effects arising from the 
tagging $B$ meson decay~\cite{tagside}.

Results from an analysis using a similar technique have been 
published by BaBar~\cite{babar_partial},
and both BaBar~\cite{babar_full} and Belle~\cite{belle_full}
have published results using full reconstruction.
This measurement is based on a $140 \ {\rm fb}^{-1}$ data sample,
which contains 152 million $B\overline{B}$ pairs, 
collected  with the Belle detector at the KEKB asymmetric-energy
$e^+e^-$ ($3.5 \ {\rm GeV}$ on $8 \ {\rm GeV}$) collider~\cite{KEKB}.
KEKB operates at the $\Upsilon(4S)$ resonance 
($\sqrt{s} = 10.58 \ {\rm GeV}$) with a peak luminosity that exceeds
$1.5 \times 10^{34}~{\rm cm}^{-2}{\rm s}^{-1}$.

At KEKB, the $\Upsilon(4S)$ is produced
with a Lorentz boost of $\beta\gamma=0.425$ antiparallel to the positron beam
direction ($z$).
Since the $B^0$ and $\bar{B}^0$ mesons are approximately at 
rest in the $\Upsilon(4S)$ center-of-mass system (cms),
$\Delta t$ can be determined from the displacement in $z$ 
between the two $B$ meson decay vertices:
\begin{equation}
  \label{eq:dt_ideal}
  \Delta t \simeq (z_{\rm sig} - z_{\rm tag})/\beta\gamma c \equiv \Delta z/\beta\gamma c.
\end{equation}
The vertex positions $z_{\rm sig}$ and $z_{\rm tag}$
are obtained independently from the fast pion and tagging lepton, respectively.

The Belle detector is a large-solid-angle magnetic spectrometer that
consists of a three-layer silicon vertex detector (SVD),
a 50-layer central drift chamber (CDC), 
an array of aerogel threshold \v{C}erenkov counters (ACC), 
a barrel-like arrangement of time-of-flight scintillation counters (TOF), 
and an electromagnetic calorimeter comprised of CsI(Tl) crystals (ECL) 
located inside a superconducting solenoid coil that provides a 
$1.5 \ {\rm T}$ magnetic field.  
An iron flux-return located outside of the coil is instrumented 
to detect $K_L^0$ mesons and to identify muons (KLM).  
The detector is described in detail elsewhere~\cite{Belle}.

Candidate events are selected by requiring the presence of
fast pion and slow pion candidates.
In order to obtain accurate vertex position determinations,
fast pion candidates are 
required to originate from the interaction point,
to have associated hits in the SVD,
and to have a polar angle in the laboratory frame in the range
$30^\circ < \theta_{\rm lab} < 135^\circ$.
The vertex positions are obtained by fits 
of the candidate tracks with the run-dependent interaction point profile,
which is smeared to account for the $B$ meson decay length.
Fast pion candidates are required to be inconsistent 
with either lepton hypothesis (see below), 
and also with a kaon hypothesis,
based on information from the CDC, TOF and ACC. 
A requirement on the fast pion cms momentum of
$1.83 \ {\rm GeV}/c < p_{\pi_f} < 2.43 \ {\rm GeV}/c$ is made;
this range includes both signal and sideband regions (defined below).
Slow pion candidates are required to have cms momentum in the range
$0.05 \ {\rm GeV}/c < p_{\pi_s} < 0.30 \ {\rm GeV}/c$.
No requirement is made on particle identification for slow pions,
and since they are not used for vertexing only a loose requirement 
that they originate from the interaction point is made.

In order to reduce background from continuum 
$e^+e^- \to q\bar{q} \ (q = u,d,s,c)$ processes,
we require the presence of a high-momentum lepton in the event.
Tagging lepton candidates are required to be positively identified
either as electrons, on the basis of information from the CDC, ECL and ACC, 
or as muons, on the basis of information from the CDC and the KLM.
They are required to have cms momentum in the range
$1.2 \ {\rm GeV}/c < p_{l_{\rm tag}} < 2.3 \ {\rm GeV}/c$,
and to have a cms angle with the fast pion candidate which satisfies
$-0.75 < \cdfl$.
The lower bound on the momentum and the requirement on the angle 
also reduce, to a negligible level,
the contribution of leptons produced from semi-leptonic
decays of the unreconstructed $D$ mesons in the $B^0 \to D^{*\mp}\pi^\pm$ decay chain.
No other tagging lepton candidate with momentum greater than 
$1.0 \ {\rm GeV}/c$ is allowed in the event to reduce the 
mistagging probability, and also to reduce the contribution
from leptonic charmonium decays.
Identical vertexing requirements to those for fast pion candidates 
are made in order to obtain an accurate $z_{\rm tag}$ position.
To further suppress the small remaining continuum background,
we impose a loose requirement on the ratio of 
the second to zeroth Fox-Wolfram~\cite{fw} moments, $R_2 < 0.6$.

Signal events are distinguished from background using three kinematic
variables, which are approximately independent for signal.
These are denoted by $p_{\pi_f}$, $\cdfs$ and $\cts$.
For signal, the fast pion cms momentum, $p_{\pi_f}$,
has a uniform distribution, smeared by the experimental resolution,
as the fast pion is monoenergetic in the $B$ rest frame.
The cosine of the angle between the fast pion direction and the 
opposite of the slow pion direction in the cms, $\cdfs$,
peaks sharply at $+1$ for signal,
as the slow pion follows the $D^*$ direction, 
due to the small energy released in the $D^*$ decay.
The angle between the slow pion direction and the opposite of the 
$B$ direction in the $D^*$ rest frame, $\cts$,
has a distribution proportional to $\ctssq$ for signal events,
as the $B$ decay is a pseudoscalar to pseudoscalar vector transition.
Since the $D^*$ is not fully reconstructed, 
$\cts$ is calculated using kinematic constraints, 
and the background can populate the unphysical region $\left|\cts\right|>1$.

We select candidates which satisfy
$1.83 \ {\rm GeV}/c < p_{\pi_f} < 2.43 \ {\rm GeV}/c$, 
$0.850 < \cdfs < 1.000$ and $-1.35 < \cts  < 1.80$.
In the cases where more than one candidate satisfies these criteria,
we select the one with the largest value of $\cdfs$.
We further define signal regions in $p_{\pi_f}$ and $\cdfs$ as 
$2.13 \ {\rm GeV}/c < p_{\pi_f} < 2.43 \ {\rm GeV}/c$, $0.925 < \cdfs < 1.000$,
and two regions in $\cts$:
$-1.00 < \cts  < -0.30$ and $+0.40 < \cts  < +1.10$.

Background events are separated into three categories:
$D^{*\mp}\rho^{\pm}$, which is kinematically similar to the signal; 
correlated background, in which the slow pion originates from the decay of 
a $D^*$ which in turn originates from the decay of the same $B$ 
as the fast pion candidate ({\it e.g.} $D^{**}\pi$);
uncorrelated background, which includes everything else
({\it e.g.} continuum processes, $D\pi$).
The kinematic distributions of the background categories and the signal
are determined from a large Monte Carlo (MC) sample,
corresponding to three times the integrated luminosity of our data sample,
in which the branching fractions of the signal and major background 
sources are reweighted according to the 
most recent knowledge~\cite{pdg,kuzmin}.
We also use this MC sample for various tests of the analysis algorithms.

\begin{figure}[htb]
  \includegraphics[width=0.48\textwidth]{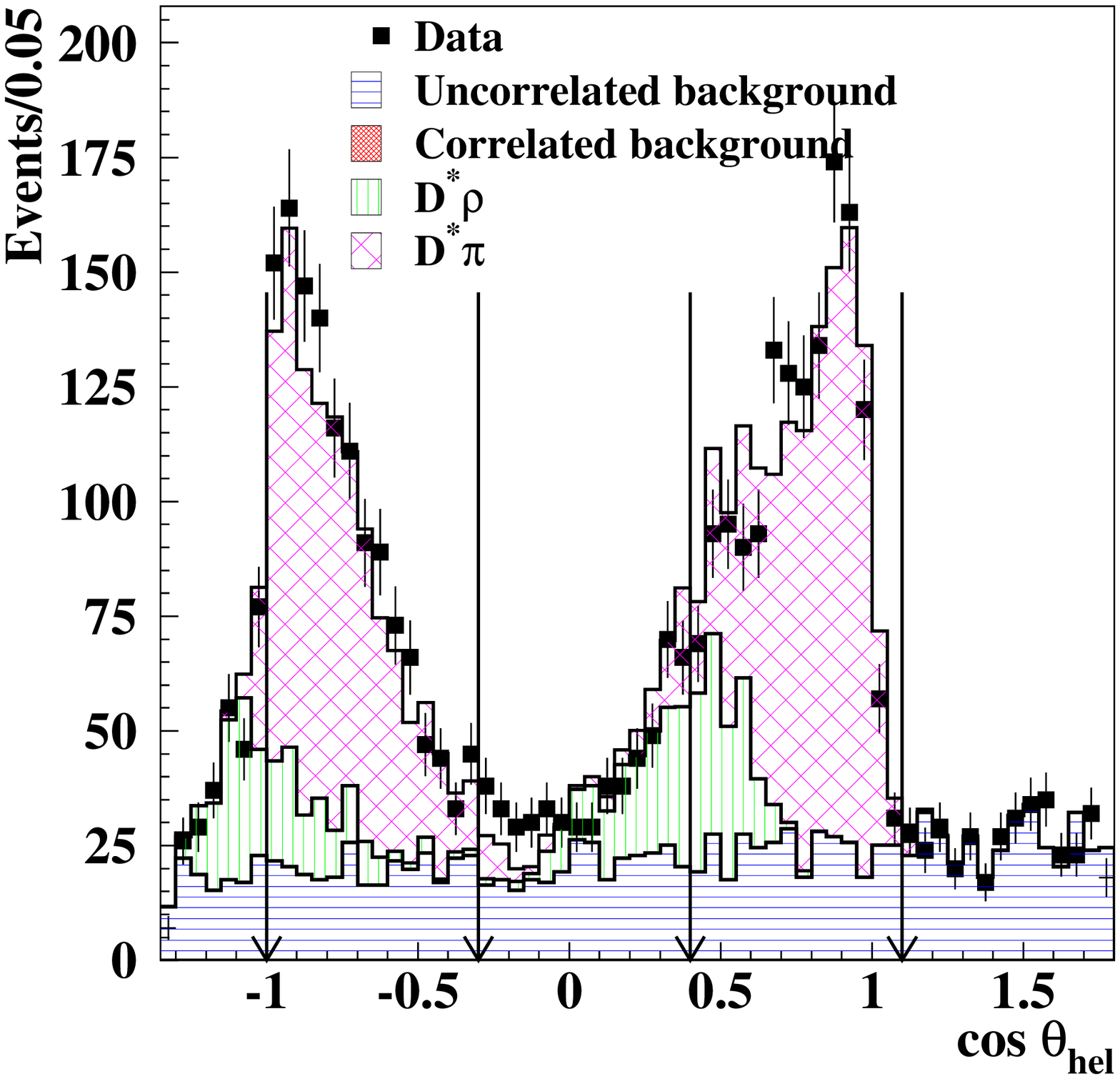}   
  \includegraphics[width=0.48\textwidth]{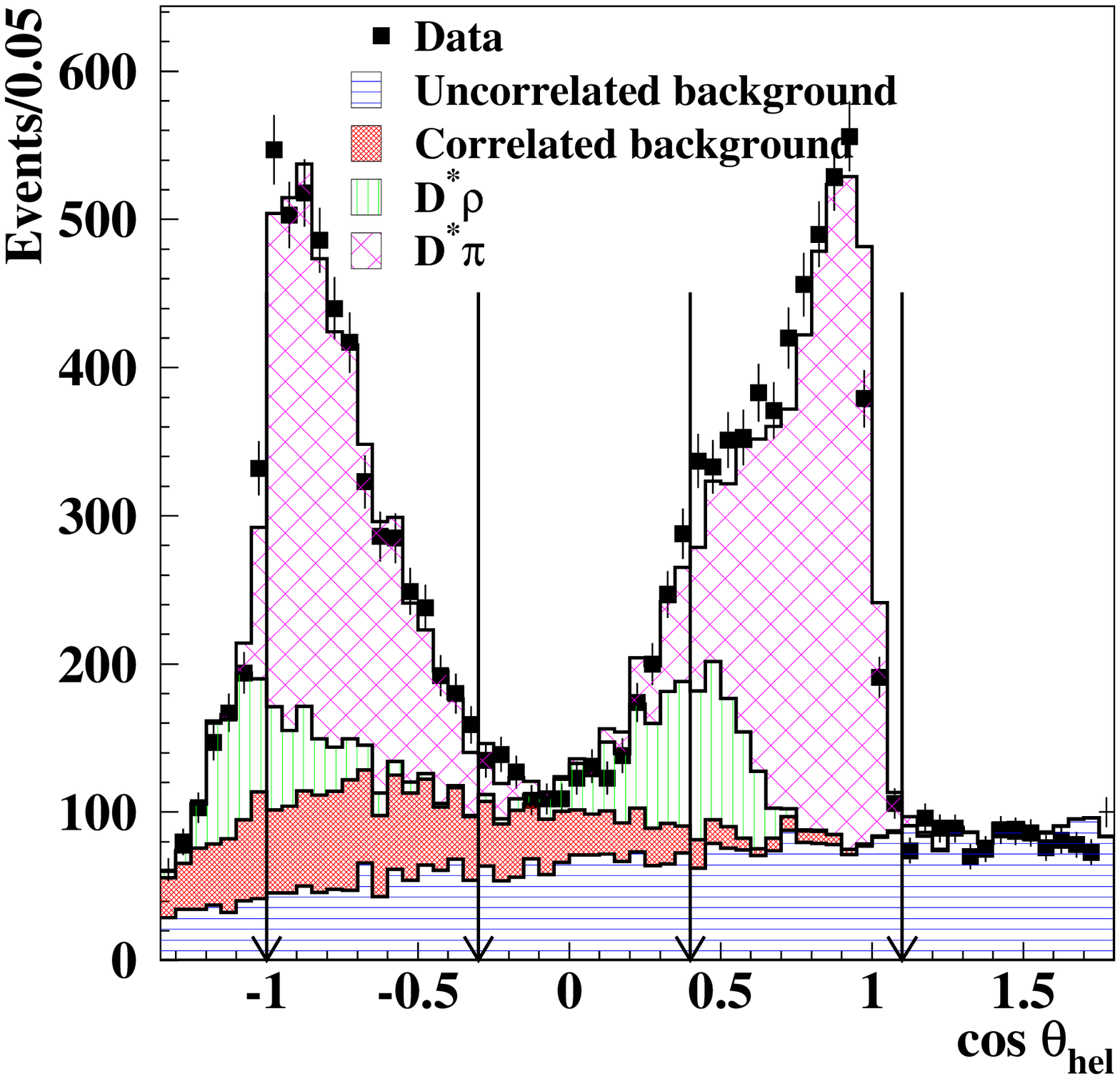}
  \caption{
    \label{fig:kin_fit}
    Results of the kinematic fits to 
    (left) same flavour and (right) opposite flavour $D^*\pi$ candidates,
    in the $p_{\pi_f}$ and $\cdfs$ signal regions,
    projected onto the $\cts$ axis.
    The horizontally hatched area represents the uncorrelated background,
    the filled area represents the correlated background,
    the vertically hatched area represents the $D^*\rho$ component
    and the cross-hatched area is the signal.
    The arrows show the edges of the signal regions.
  }
\end{figure}

Event-by-event signal fractions are determined from 
binned maximum likelihood fits to the three-dimensional 
kinematic  distributions
(6 bins of $p_{\pi_f}$ $\times$ 6 bins of $\cdfs$ $\times$ 9 bins of $\cts$).
Separate fits are performed for same flavour (SF) events,
in which the fast pion and the tagging lepton have the same charge,
and opposite flavour (OF) events,
in which the fast pion and the tagging lepton have opposite charges.
MC studies show that there is little correlated background 
in the SF sample
(since most of the correlated background is found to 
originate from charged $B$ decays~\cite{kuzmin},
that, in the limit of perfect tagging, contribute only to the OF sample), 
so this contribution is constrained to be zero.
For the determination of the signal fractions,
only events in which the fast and slow pion have opposite charges
(right sign events) are considered.
(Wrong sign events, in which the two pions have the same charge,
are used to model the uncorrelated background, as described below.)
The results of these fits, projected onto the $\cts$ axis for
events in the signal regions of $p_{\pi_f}$ and $\cdfs$
are shown in Fig.~\ref{fig:kin_fit},
and summarised in Table~\ref{tab:kin_fit}.

\begin{table}[htb]
  \caption{
    \label{tab:kin_fit}
    Summary of the results of the three-dimensional fits to 
    kinematic variables.
    The numbers of events given for each category 
    are those extrapolated to inside 
    the signal regions in all three variables.
  }
  \begin{center}
    \begin{tabular*}{\textwidth}
      {
        c
        @{\extracolsep{\fill}}
        r@{$\,\pm\,$\extracolsep{0pt}}l
        @{\extracolsep{\fill}}
        r@{$\,\pm\,$\extracolsep{0pt}}l
        @{\extracolsep{\fill}}
        r@{$\,\pm\,$\extracolsep{0pt}}l
        @{\extracolsep{\fill}}
        r@{$\,\pm\,$\extracolsep{0pt}}l
        @{\extracolsep{\fill}}
        r@{$\,\pm\,$\extracolsep{0pt}}l
        @{\extracolsep{\fill}}
      }
      \hline
      Mode & 
      \multicolumn{2}{c} {Data}   & 
      \multicolumn{2}{c} {$D^*\pi$} & 
      \multicolumn{2}{c} {$D^*\rho$} & 
      \multicolumn{2}{c} {Corr. bkgd} & 
      \multicolumn{2}{c} {Unco. bkgd} \\
      \hline
      SF &  2823 &  53 & 1908 &  60 & 311 & 12 & \multicolumn{2}{c} {0 (fixed)}   &  637.0 &  8.9 \\
      OF & 10078 & 100 & 6414 & 109 & 777 & 21 & 928 & 20 & 1836 & 18 \\
      \hline
    \end{tabular*}
  \end{center}
\end{table}

In order to measure the $CP$ violation parameters in the $D^*\pi$ sample,
we perform an unbinned fit to the 
SF and OF right sign candidates which are in the signal regions
of all three kinematic variables.
We minimize the quantity $-2\ln {\cal L} = -2 \sum_i \ln {\cal L}_i$,
where 
\begin{equation}
  \label{eq:likelihood}
  {\cal L}_i = 
  f_{D^*\pi} P_{D^*\pi} + f_{D^*\rho} P_{D^*\rho} +
  f_{\rm unco} P_{\rm unco} + f_{\rm corr} P_{\rm corr}
\end{equation}

The event-by-event signal and background fractions
(the $f$ terms) are taken from the results of the kinematic fits.
Each $P$ term contains an underlying physics 
probability density function (PDF),
with experimental effects taken into account. 
For $D^*\pi$ and $D^*\rho$, the underlying PDF is given by Eq.~\ref{eq:pdf},
where for $D^*\rho$ the terms $S^\pm$ are effective parameters
averaged over the helicity states~\cite{dstarrho}
that are constrained to be zero.
The underlying PDF for correlated background contains
some fraction from neutral $B$ decays 
(given by Eq.~\ref{eq:pdf} with $S^\pm = 0$),
and the remainder from charged $B$ decays
(for which the PDF is $\frac{1}{2\tau_{B^+}} e^{-\left| \Delta t \right| / \tau_{B^+}}$,
where $\tau_{B^+}$ is the lifetime of the charged $B$ meson).
The underlying PDF for uncorrelated background contains both
neutral and charged $B$ components, 
with the remainder from continuum 
$e^+e^- \to q\bar{q} \ (q = u,d,s,c)$ processes.
The underlying continuum PDF is modelled with two components; 
one with negligible lifetime,  and the other with a finite lifetime.
The sideband parameters are determined from data sidebands, 
as described later.

As mentioned above, experimental effects need to be taken into
account to obtain the $P$ terms of Eq.~\ref{eq:likelihood}.
Mistagging is taken into account using
\begin{equation}
  \begin{array}{ccc}
    \label{eq:exp_pdf}
    P( l_{\rm tag}^-, \pi_f^\pm) & = & 
    ( 1 - w_- )  P(B^{0} \to D^{*\mp} \pi^\pm) + 
    w_+ P(\bar{B}^{0} \to D^{*\mp} \pi^\pm),
    \\
    P( l_{\rm tag}^+, \pi_f^\pm) & = & 
    ( 1 - w_+ )  P(\bar{B}^{0} \to D^{*\mp} \pi^\pm) + 
    w_- P(B^{0} \to D^{*\mp} \pi^\pm),
    \\
  \end{array}
\end{equation}
where $w_+$ and $w_-$ are respectively the probabilities to incorrectly 
measure the flavour of tagging $B^0$ and $\bar{B}^0$ mesons 
(wrong tag fractions),
and are determined from the data as free parameters in the fit for $S^\pm$.

The time difference $\Delta t$ is related to the measured quantity $\Delta z$
as described in Eq.~\ref{eq:dt_ideal}, 
with additional consideration to possible offsets 
in the mean values of $\Delta z$,
\begin{equation}
  \label{eq:dt_offset}
  \Delta t \longrightarrow \Delta t + \epsilon_{\Delta t} \simeq \left( \Delta z + \epsilon_{\Delta z} \right) / \beta\gamma c.
\end{equation}
It is essential to allow non-zero values of $\epsilon$ since a 
small bias can mimic the effect of $CP$ violation:
\begin{equation}
  \begin{array}{l@{\hspace{50mm}}cr}
    \multicolumn{2}{l}{
      C \cos (\Delta m \Delta t) \longrightarrow
      C \cos (\Delta m ( \Delta t + \epsilon_{\Delta t} )) \simeq
    } \\
    &
    \multicolumn{2}{r}{
      C \cos (\Delta m \Delta t) - \Delta m \epsilon_{\Delta t} \sin (\Delta m \Delta t).
    }
  \end{array}
\end{equation}
Thus a bias as small as $\epsilon_{\Delta z} \sim 1 \ \mu{\rm m}$ can lead to 
sine-like terms as large as $0.01$, 
comparable to the expected size of the $CP$ violation effect.
We allow separate offsets for each particular combination of fast pion and 
tagging lepton charge.
We also apply a small correction to each measured vertex position
to correct for a known bias due to relative misalignment of the SVD and CDC.
This correction is dependent on the track charge, 
momentum and polar angle, measured in the laboratory frame.
It is obtained by comparing the vertex positions 
calculated with the alignment constants used in the data,
to those obtained with an improved set of alignment constants~\cite{dzb}.

Resolution effects are taken into account in a way similar to our
other time-dependent analyses~\cite{belle_sin2phi1}.
The algorithm includes components related to detector resolution
and kinematic smearing.
Since, for correctly tagged signal events,
both the fast pion and the tagging lepton 
originate directly from $B$ meson decays,
we do not include any additional smearing due to non-primary tracks.
Incorrectly tagged events can originate from secondary tracks - 
since the wrong tag fractions are small, 
we neglect their effect on the resolution function.
The effect of the approximation that the $B$ mesons are at rest
in the cms in Eq.~\ref{eq:dt_ideal} is taken into account~\cite{vertex_nim}.
We use a slightly modified algorithm to describe the detector resolution,
in order to precisely describe the observed behaviour for 
single track vertices. 
The resolution for each track is described by the sum of 
three Gaussian components, with a common mean of zero,
and widths which are given by the measured vertex error for each track
multiplied by different scale factors.

We measure the five parameters of the detector resolution function 
(three scale factors and two parameters giving the relative normalizations
of the Gaussians)
using $\jpsi \to \mu^+\mu^-$ candidates.
These are selected using similar criteria to those for $D^*\pi$,
except that both tracks are required to be identified as muons, 
and their invariant mass is required to be consistent with that of the $\jpsi$.
Vertex positions are obtained independently for each track,
in the same way as described above.
Then $\Delta z = z_{\mu^+} - z_{\mu^-}$ describes the detector resolution,
which is the convolution of the two vertex resolutions,
since for $\jpsi \to \mu^+\mu^-$ the underlying PDF is a delta function.

\begin{figure}[!htb]
  \begin{center}
    \includegraphics[width=0.38\textwidth]{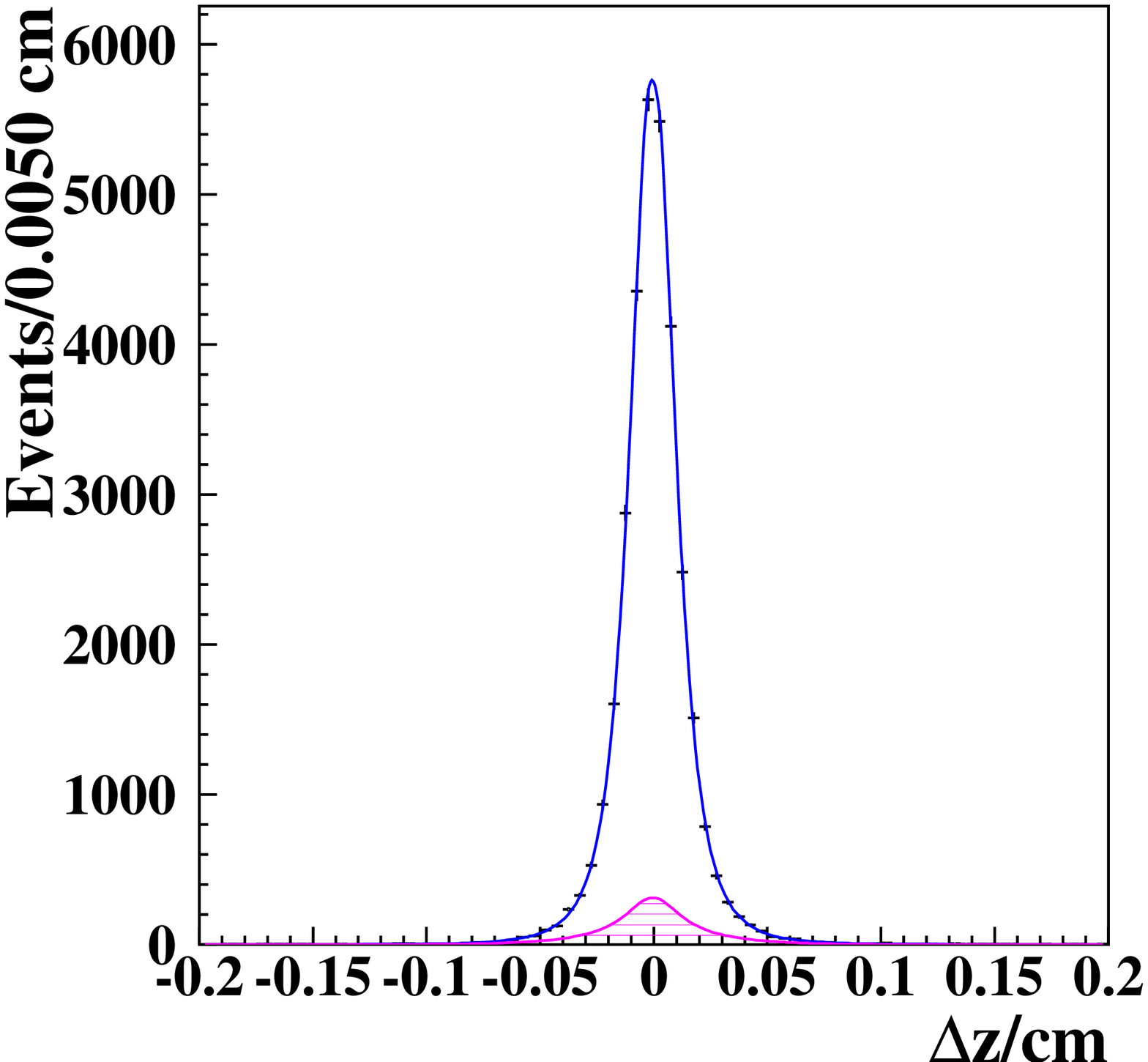}   
    \includegraphics[width=0.38\textwidth]{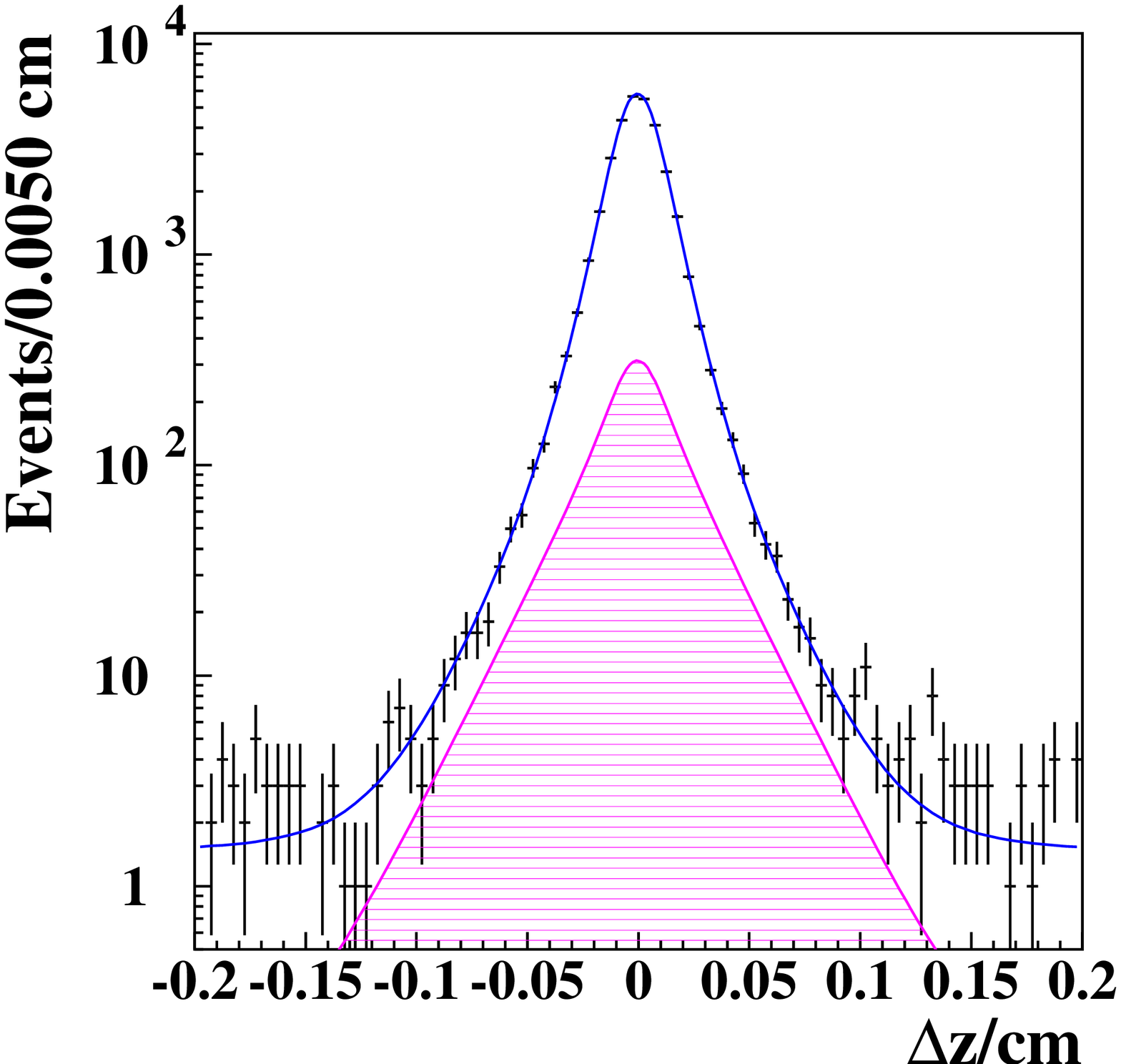}   
  \end{center}
  \caption{
    \label{fig:fit_jpsi}
    Result of the resolution parameter extraction procedure,
    for $\jpsi \to \mu^+\mu^-$ candidates selected from data,
    shown with both (left) linear and (right) logarithmic ordinate scales.
    The data points show the $\Delta z$ distribution for candidates
    in the signal region; the curve shows the result of the fit.
    The horizontally hatched area indicates the background contribution.
  }
\end{figure}

We perform an unbinned maximum likelihood fit using events 
in the $\jpsi$ signal region in the di-muon invariant mass,
and using sideband regions to determine the shape of the background 
under the peak.
The underlying $\Delta z$ PDF of the background is parametrized in the same way 
as that used for continuum, described above.
We also take a possible offset into account,
so that there are in total eight free parameters in this fit
(five describing the resolution function,
two describing the background, and one $\Delta z$ offset),
the results of which are shown in Fig.~\ref{fig:fit_jpsi}.

The free parameters in the background PDFs $P_{\rm unco}$ and $P_{\rm corr}$ 
are determined using data from sideband regions.
To measure the uncorrelated background shape,
we use wrong sign events (where the slow and fast pions have the same sign)
in the signal region.
Since the number of events in this region is quite limited,
we neglect the contribution to the uncorrelated background
from $B\bar{B}$ events,
so that the underlying PDF contains only the part due to continuum,
which has two free parameters.
We perform separate fits for the same flavour and opposite flavour 
wrong sign candidates.

To obtain the correlated background parameters,
a simultaneous fit is carried out to right and wrong sign events 
in a sideband region of
$1.83 \ {\rm GeV}/c < p_{\pi_f} < 2.03 \ {\rm GeV}/c$ and $-1.0 < \cts < -0.3$ 
(in the signal region of $\cdfs$).
This sideband region is dominated by correlated and 
uncorrelated backgrounds \textemdash{} the contributions from $D^*\pi$ and $D^*\rho$ 
are found to be small in MC \textemdash{} and the wrong sign events 
come from uncorrelated background only.
The uncorrelated background is treated as before,
whilst the correlated background contains contributions from
$B^+B^-$ and $B^0\bar{B}^0$, as described above.
Since the correlated background contribution is constrained to be
zero for same flavour events, this fit is performed for opposite flavour only.

In order to test our fit procedure,
we first constrain $S^+$ and $S^-$ to be zero and perform a fit in
which $\tau_{B^0}$ and $\Delta m$ 
(as well as two wrong tag fractions and four offsets) are free parameters.
We obtain values of $\tau_{B^0} = 1.523 \pm 0.020 \ {\rm ps}$
and $\Delta m = 0.512 \pm 0.010 \ {\rm ps}^{-1}$,
where the errors are statistical only.
These are in excellent agreement with the current world averages of
$\tau_{B^0} = 1.536 \pm 0.014 \ {\rm ps}$ and 
$\Delta m = 0.502 \pm 0.007 \ {\rm ps}^{-1}$~\cite{pdg}.
Reasonable agreement with the input values is also obtained in MC.
Furthermore, fits to the MC with $S^\pm$ floated give results 
consistent with zero, as expected.

\begin{figure}[htb]
  \begin{center}
    \includegraphics[width=0.38\textwidth]{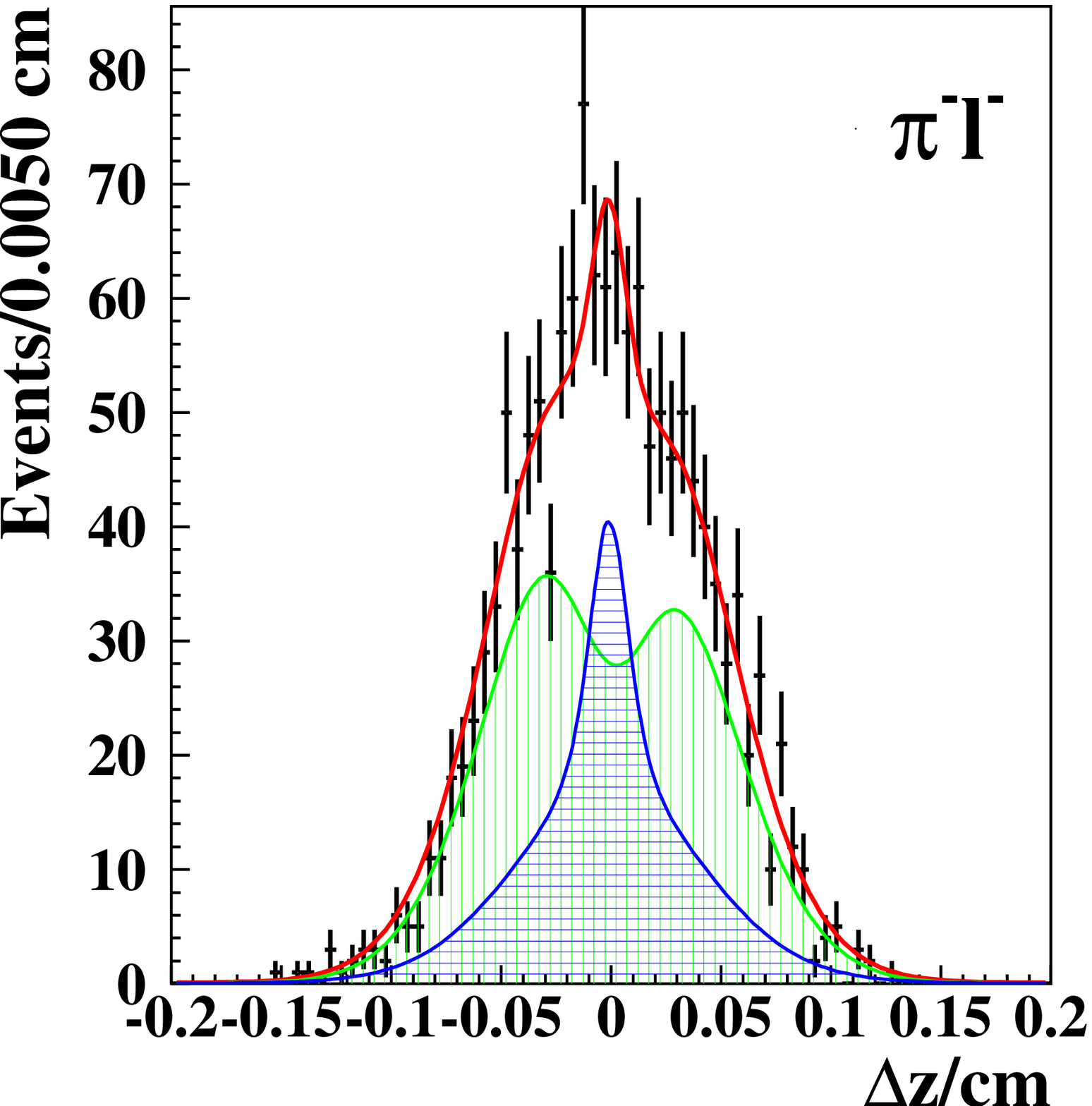}
    \includegraphics[width=0.38\textwidth]{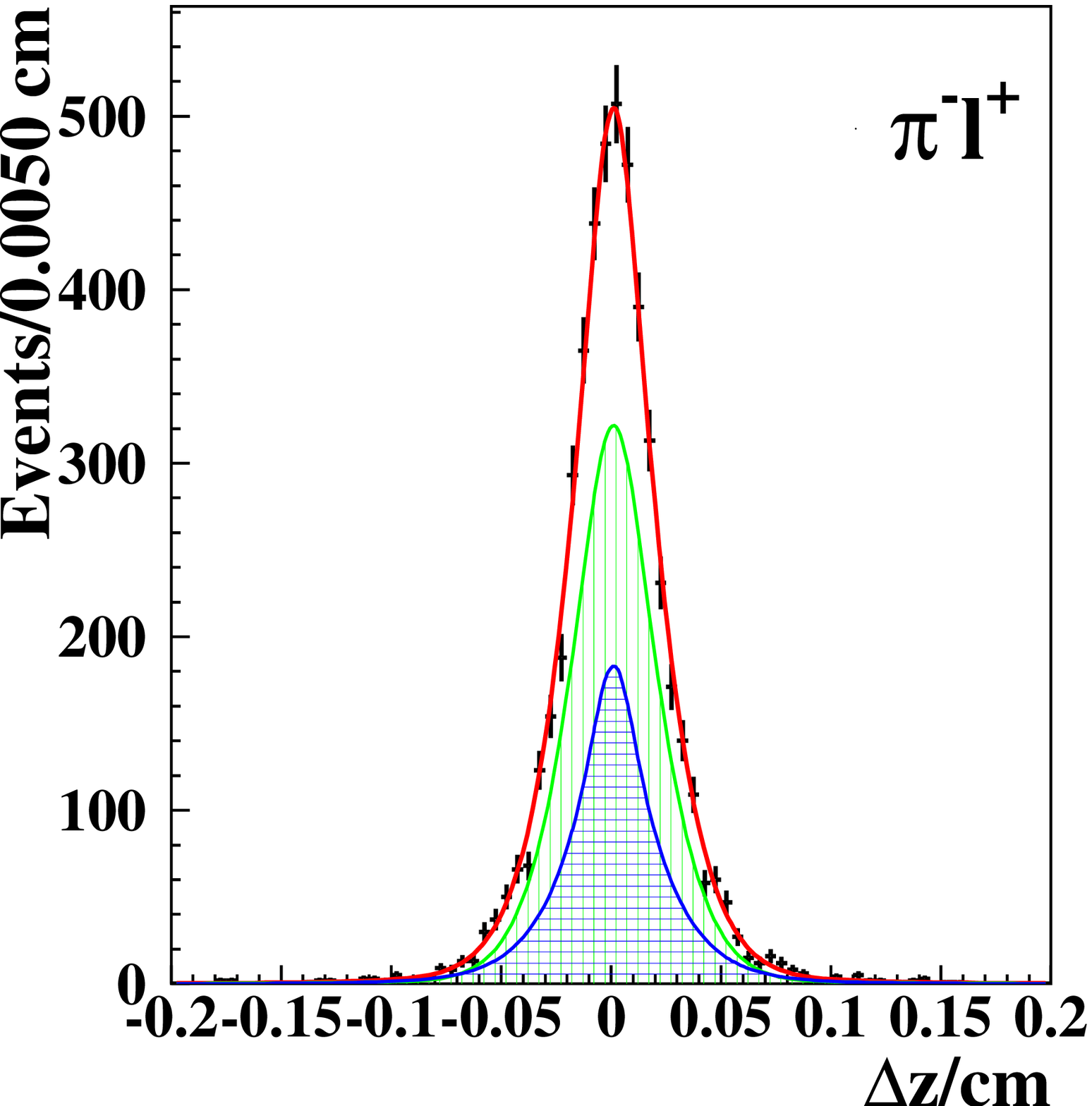} \\
    \includegraphics[width=0.38\textwidth]{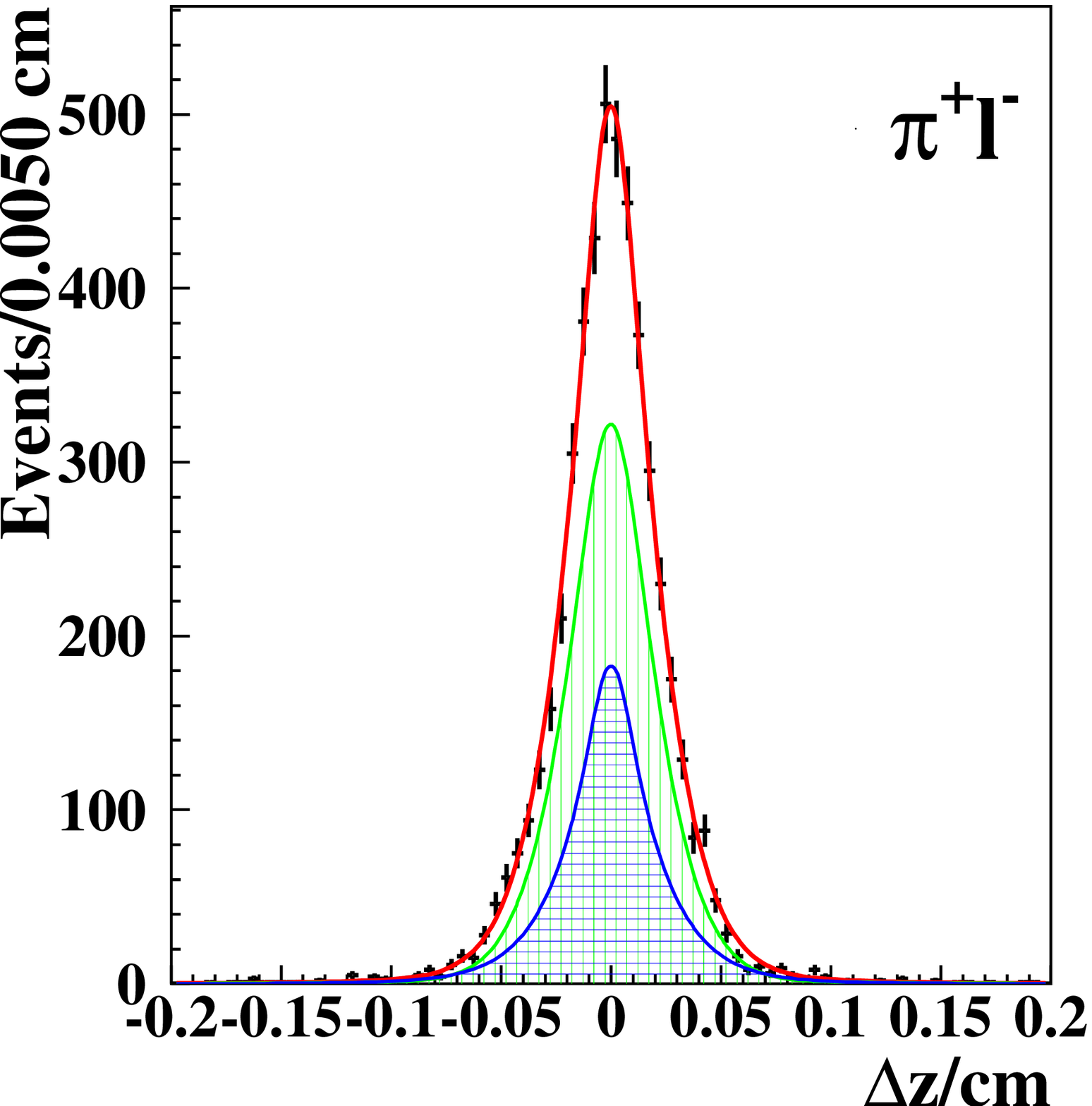}
    \includegraphics[width=0.38\textwidth]{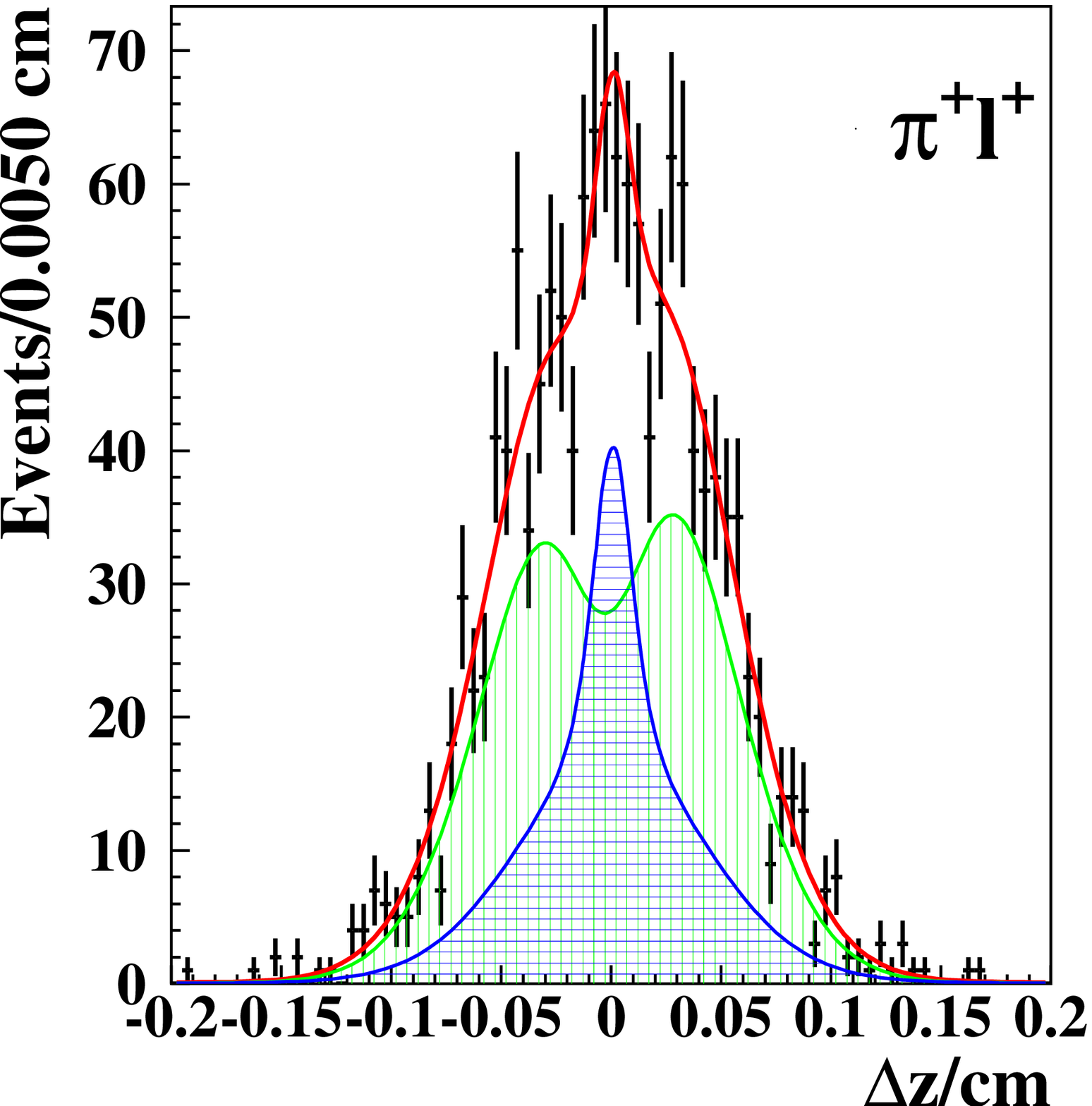}

  \end{center}
  \caption{
    \label{fig:myfit}
    Results of the fit to obtain $S^+$ and $S^-$.
    The fit result is superimposed on the data.
    The signal component is shown as the vertically hatched area.
    The horizontally hatched area indicates the background contribution.
  }
\end{figure}

\begin{figure}[htb]
  \begin{center}
    \includegraphics[width=0.38\textwidth]{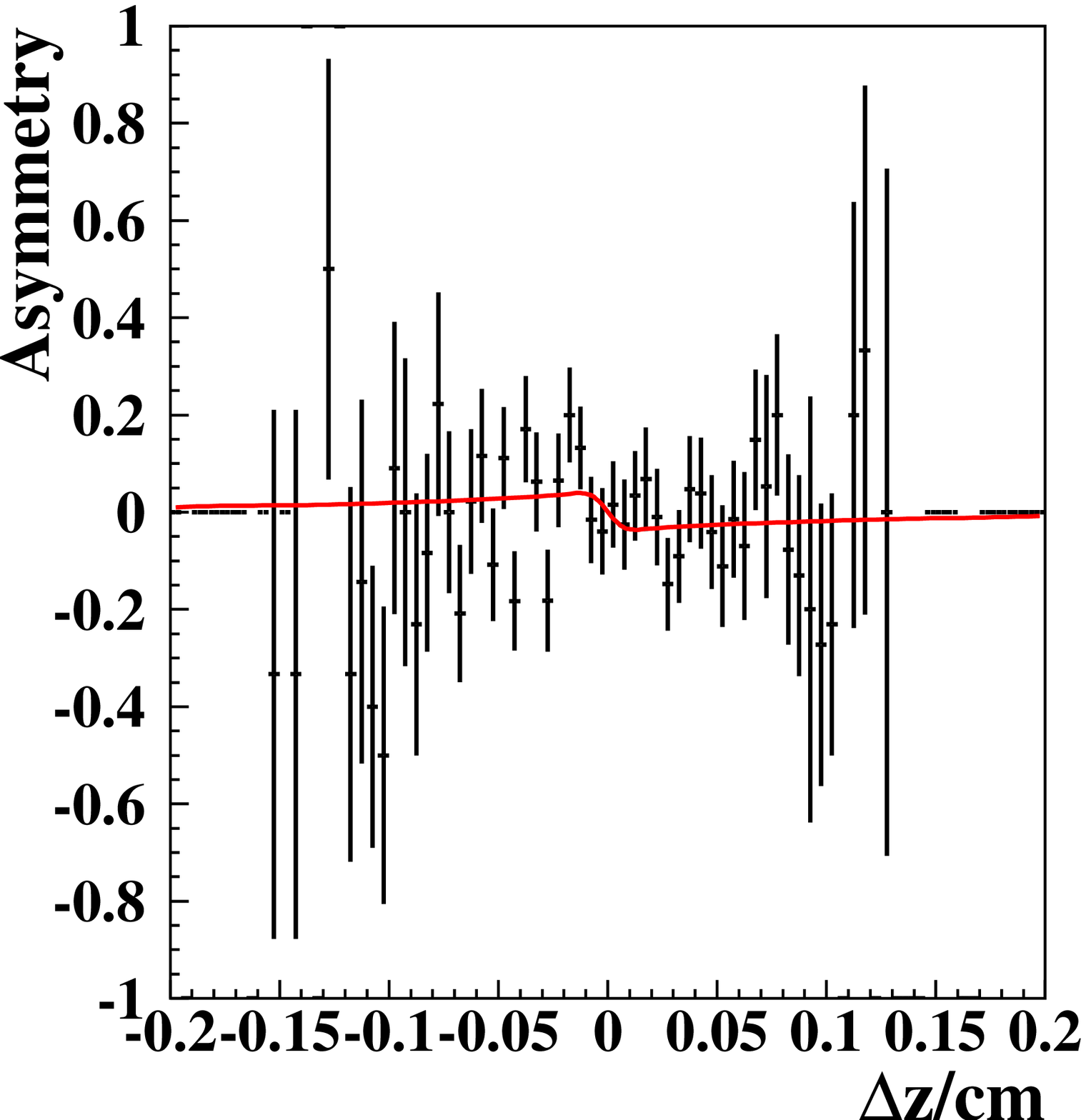}
    \includegraphics[width=0.38\textwidth]{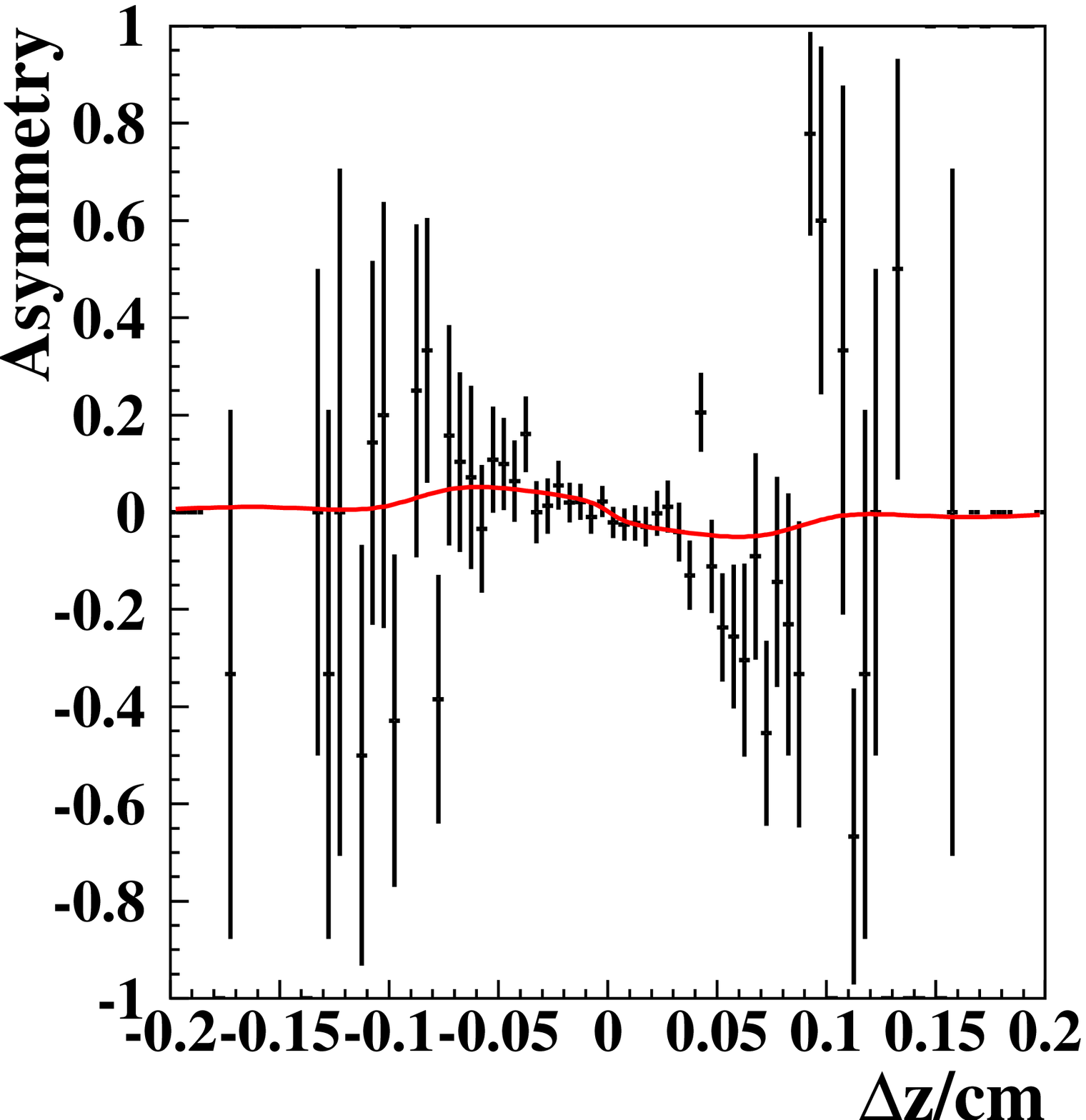}
  \end{center}
  \caption{
    \label{fig:myfit_asp}
    Results of the fit to obtain $S^+$ and $S^-$,
    shown as asymmetries in the 
    (left) same flavour events and
    (right) opposite flavour events.
    The fit result is superimposed on the data.
  }
\end{figure}

To extract the $CP$ violation parameters we fix $\tau_{B^0}$ and $\Delta m$ at their 
world average values,
and fit with $S^+$, $S^-$, two wrong tag fractions and four offsets
as free parameters.
We obtain
$S^+ = 0.035 \pm 0.041$ and $S^- = 0.025 \pm 0.041$
where the errors are statistical only.
The wrong tag fractions are $w_- = (5.1 \pm 0.5)\%$ and $w_+ = (5.2 \pm 0.5)\%$.
The results are shown in Fig.~\ref{fig:myfit}.
To further illustrate the CP violation effect,
we define asymmetries in the 
same flavour events (${\cal A}^{\rm SF}$) 
and in the opposite flavour events (${\cal A}^{\rm OF}$), as 
\begin{eqnarray}
  {\cal A}^{\rm SF} & = &
  \left( N_{\pi^-l^-}(\Delta z) - N_{\pi^+l^+}(\Delta z) \right) / 
  \left( N_{\pi^-l^-}(\Delta z) + N_{\pi^+l^+}(\Delta z) \right) \ {\rm and} \nonumber \\
  {\cal A}^{\rm OF} & = &
  \left( N_{\pi^+l^-}(\Delta z) - N_{\pi^-l^+}(\Delta z) \right) / 
  \left( N_{\pi^+l^-}(\Delta z) + N_{\pi^-l^+}(\Delta z) \right), \nonumber
\end{eqnarray}
where the $N$ values indicate the number of events 
for each combination of fast pion and tag lepton charge.
These are shown in Fig.~\ref{fig:myfit_asp}.
Note that due to the relative contributions of the
sine terms in Eq.~\ref{eq:pdf}
vertex biases ({\it i.e.} non-zero offsets)
can induce an opposite flavour asymmetry,
whereas the same flavour asymmetry is more robust.

As mentioned above,
vertex biases may lead to a large systematic error on $S^\pm$,
so we have introduced offsets
to make our analysis relatively insensitive to such biases.
These additional free parameters cause an increase in the 
statistical error of about $20\%$.
To test the robustness of our algorithm,
we use a control sample of lepton tagged, 
partially reconstructed $D^*l\nu$ decays~\cite{fronga}.
Events are selected from the same data sample of $140 \ {\rm fb}^{-1}$,
and as for $D^*\pi$ we also select events from MC.
Since semileptonic $B$ decays are flavour specific,
the time-dependent decay rates are given by Eq.~\ref{eq:pdf}
with $S^+$ and $S^-$ equal to zero.
Non-zero values can be caused by vertexing, or other, biases.
In order to select a sample with larger statistics than,
and similar kinematics to, our signal,
we select $D^*l\nu$ candidates with lepton cms momentum in the range
$1.80 \ {\rm GeV}/c < p_{l_{\rm sig}} < 2.30 \ {\rm GeV}/c$
and missing mass squared~\cite{fronga} in the range
$-4.0 \ {\rm GeV}^2/c^4 < M_\nu^2 < 1.0 \ {\rm GeV}^2/c^4$;
signal regions in these two variables are defined as
$1.95 \ {\rm GeV}/c <  p_{l_{\rm sig}} < 2.30 \ {\rm GeV}/c$ and
$-1.0 \ {\rm GeV}^2/c^4 < M_\nu^2 < 1.0 \ {\rm GeV}^2/c^4$, respectively.
The cms momentum range for the tagging lepton is chosen
to make the signal and tagging lepton candidates mutually exclusive:
$1.20 \ {\rm GeV}/c < p_{l_{\rm tag}} < 1.90 \ {\rm GeV}/c$.

The entire analysis procedure is repeated for the $D^*l\nu$ candidates.
To perform the kinematic fit, $p_{l_{\rm sig}}$ and $M_\nu^2$
are used as the discriminating variables,
and backgrounds are categorised in a similar way as for the $D^*\pi$ analysis.
For $D^*l\nu$ there is no equivalent to the $D^*\rho$ background, however.
At the end of the procedure, we obtain values of 
$S^+ = -0.007 \pm 0.024$ and $S^- = -0.019 \pm 0.024$,
consistent with the expectation of zero.
We estimate the systematic error due to vertexing biases
by repeating the fits to $D^*l\nu$ candidates without 
applying the bias correction.
The largest change in $S^\pm$ is $0.003$,
which we assign as a systematic error due to vertexing.
We also perform the same study using $D^*l\nu$ events selected 
from MC, as well as with the $D^*\pi$ events (both data and MC),
and in all cases obtain consistent results.

The results of the fits to the control samples 
validate our analysis procedure.
We have further tested our fit routine for possible fit biases,
such as could be caused by neglecting terms of $\Rdp^2$ in the fit,
by generating a number of large samples of signal Monte Carlo
with different input values of $S^+$ and $S^-$.
For reasonable input values of $\Rdp$ we find no evidence of a bias,
and assign $0.005$ as the systematic error due to possible fit bias.

Other systematic errors due to the resolution function,
the background fractions, and the background parameters
are estimated by varying the values used in the fit by $\pm 1 \sigma$.
Uncertainties in the background shapes, background fractions 
and kinematic smearing parameters,
result in errors of $0.002$, $0.003$ and $0.003$, respectively.
The uncertainty in the world average values of $\tau_{B^0}$ and $\Delta m$
results in a systematic error of $0.001$.
Allowing for effective $S^\pm$ terms of $\pm0.05$ in the $D^*\rho$ PDF
leads to a systematic error of $0.004$.
The resolution function parameters are precisely determined 
from the fit to $\jpsi \to \mu^+\mu^-$ candidates.
We also consider systematic effects due to our lack of knowledge
of the exact functional form of the resolution function:
using different parametrizations results in shifts of $S^\pm$ 
as large as $0.006$, which we assign as an additional systematic error.
Similarly, systematic effects due to differences between data and MC
in the distributions used in the kinematic fit are further investigated 
by repeating the fit using different binning.
We repeat the entire fit procedure using twice as many bins 
in each of the three discriminating variables.
Since $\cdfs$ is used in the best candidate selection,
we also repeat the algorithm without using this variable in the kinematic fit.
The largest deviation ($0.010$) is assigned as an additional systematic error.
The possible bias caused by neglecting the correlated background contribution
in same flavour events is estimated to be $0.005$ by repeating the fit
allowing this component.
We estimate the effect of the approximations inherent in the 
uncorrelated background parametrization by fitting 
a sideband region of $\cdfs$, 
which is dominated by uncorrelated background,
with a parametrization which includes contributions from $B\bar{B}$ events.
Note that this fit yields $S^+$ and $S^-$ terms which are consistent 
with zero.
We then repeat the fit to the signal candidates using the 
uncorrelated background shape thus determined,
and, as before, assign the largest shift in $S^+$ or $S^-$ ($0.010$)
as the systematic error.
The systematic errors are summarized in Table~\ref{tab:systematics}.
The total systematic error ($0.018$) is obtained by adding the 
above terms in quadrature.

\begin{table}[htb]
  \caption{
    \label{tab:systematics}
    Summary of the systematic uncertainties.
  }
  \begin{center}
    \begin{tabular*}{\textwidth}
      {
        l@{\extracolsep{\fill}}c
      }
      \hline
      Source & Uncertainty \\
      \hline
      $\tau_{B^0}$, $\Delta m$                        & $0.001$ \\
      $CP$ violation in $D^*\rho$                & $0.004$ \\
      Background shapes                       & $0.002$ \\
      Correlated background parametrization   & $0.005$ \\
      Uncorrelated background parametrization & $0.010$ \\
      Background fractions                    & $0.003$ \\
      Kinematic smearing                      & $0.003$ \\      
      Resolution parametrization              & $0.006$ \\
      Vertexing                               & $0.003$ \\
      Kinematic fit binning                   & $0.010$ \\
      \hline
      Total                                   & $0.018$ \\
      \hline
    \end{tabular*}
  \end{center}
\end{table}

In summary, we have measured $CP$ violation parameters
using partially reconstructed $B^0 \to D^{*\mp} \pi^\pm$ decays,
from a data sample of $140 \ {\rm fb}^{-1}$.
The results are
\begin{eqnarray}
 S^+ & = & 0.035 \pm 0.041 \ ({\rm stat}) \pm 0.018 \ ({\rm syst}),  \nonumber \\
 S^- & = & 0.025 \pm 0.041 \ ({\rm stat}) \pm 0.018 \ ({\rm syst}). \nonumber 
\end{eqnarray}
These are consistent with, and more precise than,
the previously published measurements of $CP$ violation parameters
in $B^0 \to D^{*\mp}\pi^\pm$ decays~\cite{babar_partial,babar_full,belle_full}.
At present there are no reliable measurements of either $\Rdp$ or $\ddp$,
and so we cannot extract $\sin(2\phi_1+\phi_3)$ from our measured values.
However, measurements of $\Rdp$ are anticipated in future,
and hence more precise measurements of $S^\pm$ 
will help constrain the Standard Model.

\section*{Acknowledgments}
We thank the KEKB group for the excellent operation of the
accelerator, the KEK cryogenics group for the efficient
operation of the solenoid, and the KEK computer group and
the National Institute of Informatics for valuable computing
and Super-SINET network support. We acknowledge support from
the Ministry of Education, Culture, Sports, Science, and
Technology of Japan and the Japan Society for the Promotion
of Science; the Australian Research Council and the
Australian Department of Education, Science and Training;
the National Science Foundation of China under contract
No.~10175071; the Department of Science and Technology of
India; the BK21 program of the Ministry of Education of
Korea and the CHEP SRC program of the Korea Science and
Engineering Foundation; the Polish State Committee for
Scientific Research under contract No.~2P03B 01324; the
Ministry of Science and Technology of the Russian Federation; 
the Ministry of Higher Education, 
Science and Technology of the Republic of Slovenia;  
the Swiss National Science Foundation; 
the National Science Council and the Ministry of Education of Taiwan; 
and the U.S. Department of Energy.


\begin{thebibliography}{99}
\bibitem{dzb}
  A small amount of data is reprocessed using a set of constants which
  is corrected for the small relative misalignment of the SVD and CDC.
  The correction function is calculated from comparisons 
  of the vertex positions for tracks between the two reprocessings.
\end{thebibliography}
\end{document}